\documentclass[floatfix,preprint,
 amsmath,amssymb,
 aps,
 prl
]{revtex4-2}
\DeclareUnicodeCharacter{2212}{-}
\usepackage{lineno}

\usepackage[justification=raggedright,singlelinecheck=false]{caption}

\usepackage{caption}
\usepackage{subcaption}
\usepackage{graphicx}
\usepackage{graphicx}
\usepackage{float}
\usepackage{bm}
\usepackage{units}
\usepackage{amsmath}
\usepackage{xfrac}

\usepackage{xcolor}
\usepackage[utf8]{inputenc}
\usepackage[T1]{fontenc}

\usepackage{amsfonts}
\usepackage{booktabs}
\usepackage{lineno}
\newcommand{\edit}[1]{\textcolor{black}{#1}}

\usepackage{hhline}
\usepackage{array}

\newcolumntype{P}[1]{>{\centering\arraybackslash}p{#1}}

\usepackage{mathtools}

\begin{document}

\title{Experimental evidence of production of directional muons from a laser-wakefield accelerator}

\newcommand{\QUB}{School of Mathematics and Physics, Queen's University of Belfast, BT7 1NN, Belfast, UK}

\newcommand{\YORK}{York Plasma Institute, University of York, School of Physics,
Engineering and Technology, York, YO10 5DD, UK.}

\newcommand{\ELINP}{Extreme
Light Infrastructure-Nuclear Physics (ELI-NP)/Horia Hulubei National Institute of Physics and Nuclear
Engineering, Bucharest-Magurele 077125, Romania}

\newcommand{\STRATH}{Department of Physics, SUPA, University of Strathclyde, Glasgow G4 0NG, UK}

\newcommand{\COCK}{The Cockcroft Institute, Sci-Tech Daresbury, Warrington WA4 4AD, UK}

\newcommand{\WEIZ}{Department of Physics of Complex Systems, Weizmann Institute of Science, 234 Herzl St., Rehovot 7610001, Israel}

\newcommand{\ICL}{The John Adams Institute for Accelerator Science, Imperial College London, London, SW7 2AZ, UK}

\newcommand{\DSTL}{DSTL, Portsdown West, Fareham, United Kingdom}

\newcommand{\PRAGUE}{Institute of Experimental and Applied Physics, Czech Technical University in Prague, Husova 240/5
110 00, Czech Republic}

\newcommand{\NUCLEAR}{Doctoral School of Physics, University of Bucharest, Atomistilor 405, 077125 Magurele, Ilfov, Romania}

% corresponding order
\newcommand{\qub}{$^1$}
\newcommand{\york}{$^2$}
\newcommand{\elinp}{$^3$}
\newcommand{\prague}{$^4$}
\newcommand{\icl}{$^5$}
\newcommand{\weiz}{$^6$}
\newcommand{\strath}{$^7$}
\newcommand{\cock}{$^8$}
\newcommand{\nuclear}{$^9$}
\newcommand{\dstl}{$^{10}$}

\author{
L.~Calvin\qub,
E.~Gerstmayr\qub,
C.~Arran\york,
L.~Tudor\elinp,
T.~Foster\qub,
K.~Fleck\qub,
B.~Bergmann\prague,
D.~Doria\elinp,
B.~Kettle\icl,
H.~Maguire\qub,
V.~Malka\elinp$^,$\weiz,
P.~Manek\prague,
S.~P.~D.~Mangles\icl,
P. McKenna\strath$^,$\cock,
R.~E.~Mihai \elinp$^,$\prague,
S.~Popa\elinp$^,$\nuclear,
C.~Ridgers\york,
J.~Sarma\qub,
P.~Smolyanskiy\prague,
R.~Wilson\strath$^,$\cock,
R. M. Deas\dstl,
~and~
G.~Sarri\qub$^{,*}$
\email{g.sarri@qub.ac.uk}
}

% list addresses
\address{\qub \QUB} %1
\address{\york \YORK} %2
\address{\elinp \ELINP} %3
\address{\prague \PRAGUE}
\address{\icl \ICL}
\address{\weiz \WEIZ}
\address{\strath \STRATH}
\address{\cock \COCK}
\address{\nuclear \NUCLEAR}
\address{\dstl \DSTL}

\begin{abstract}
We report on experimental evidence of the generation of directional muons from a laser-wakefield accelerator driven by a PW-class laser. 
The muons were generated following the interaction of a GeV-scale high-charge electron beam with a \unit[2]{cm}-thick Pb target and were detected using a Timepix3 detector placed behind a suitable shielding configuration. 
Data analysis indicates a $(99.1\pm0.5)$\% confidence of muon detection over noise, in excellent agreement with numerical modelling.
Extrapolation of the experimental setup to higher electron energies and charges suggests the potential to guide and separate from noise approximately $10^4$ muons/s onto cm$^2$-scale areas for applications using a 10 Hz PW laser. These results demonstrate the possibility of generating and transporting directional muon beams using high-power lasers and establish a foundation for the systematic application of laser-driven high-energy muon beams.
\end{abstract}
\maketitle

\newpage

\noindent\textbf{I. INTRODUCTION}\\
Muon beams are attracting considerable attention from the research community thanks to their unique properties, which make them ideal candidates for disruptive applications across a wide range of scientific and industrial areas.  These include radiography of thick and dense objects \cite{Jourde2016}, muon-catalyzed fusion
\cite{Breunlich1989}, and precision muon physics \cite{Gorringe2015}. Traditionally, muon beams are either produced using radio-frequency accelerators \cite{Doble1993,Thomason2019} or naturally sourced from cosmic rays \cite{Workman2022}. Despite some remarkable proof-of-principle applications (see, e.g., \cite{MICE_2020,Morishima2017}), a systematic and widespread exploitation of muon sources is still limited either by the large size and cost of proton-driven muon sources or by the low flux of naturally occurring muons from cosmic rays (typically of the order of 1 $\mu$/cm$^2$/minute \cite{Workman2022}).

\begin{figure}[b!]
%\centering
\includegraphics[width=1\linewidth]{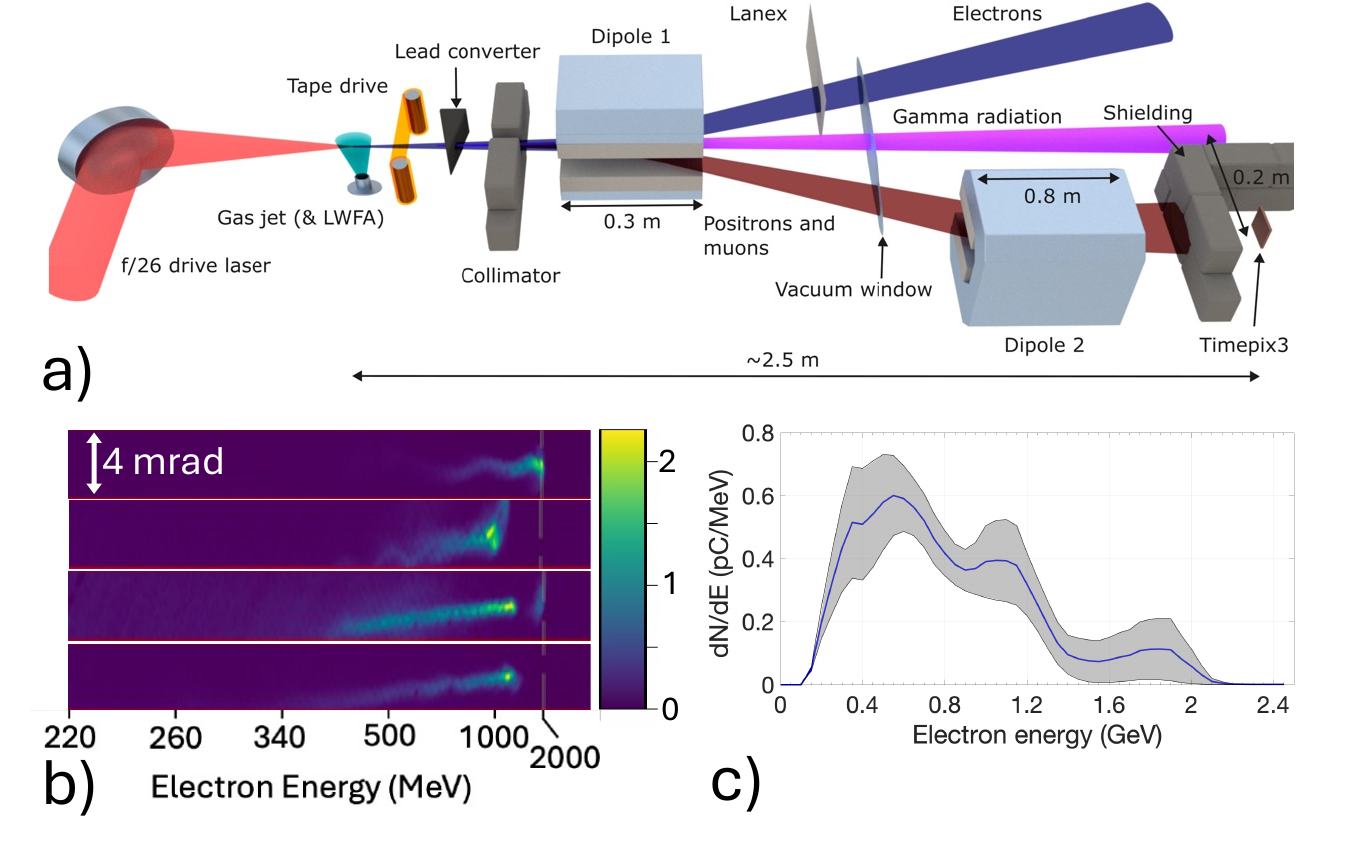}
\caption{ \textbf{a.} Sketch of the experimental setup.  
\textbf{b.} Examples of angularly resolved electron spectra measured before the insertion of the converter target. Colorbar is in arbitrary units. \textbf{c.} Angularly integrated spectrum averaged over 10 consecutive shots without converter. The solid blue line indicates the average and the grey shading the standard deviation.}
\label{Fig1}
\end{figure}

High-power lasers have been recently proposed as a viable alternative to produce sizeable muon fluxes in a relatively compact setup (see, e.g. Refs. \cite{Rao2018,Calvin2023}), with numerical work showing the possibility of achieving high-energy populations of muons with an average flux sufficient for practical applications with a turnaround as short as a few seconds to minutes \cite{Calvin2023}. \edit{Proof-of-principle experimental evidence of muon generation driven by PW-scale lasers has been independently reported \cite{Terzani2024,Zhang_muons}, where muon signal was identified by their signature decay time.}

Muons can be generated during the interaction of GeV-scale laser-wakefield accelerated electron beams with thick high-Z converters through several channels: Bethe-Heitler pair production \cite{Tsai1974}, trident processes \cite{Baier2008}, pion decay \cite{Titov2009}, and kaon decay \cite{Rosner_2001}. At GeV energies, the high mass of kaons makes its production comparatively unlikely. While pion-mediated muon production has the highest cross-section \cite{Schadmand2006}, the wide-angle emission of charged pions, combined with the large cross sections of pion scattering and absorption in heavy nuclei, results in generally not collimated muon fluxes \cite{Titov2009}. For sufficiently thick targets, Bethe-Heitler pair production dominates over direct electroproduction (trident) processes \cite{Baier2008,Titov2009}, making it the preferable mechanism for generating sizeable, directional muon fluxes from a laser-wakefield accelerator \cite{Calvin2023,Terzani2024,Titov2009}.
Bethe-Heitler pair production has been used to generate laser-driven ultra-relativistic positron beams (see, e.g., Refs. \cite{Streeter2024, Sarri2015, Sarri_ncomm_2015,Chen_2010,Li_2019}). However, the much higher rest mass of muons results in significantly lower yields compared to electron-positron pairs ($N_{\mu}/N_{e}\propto (m_e/m_\mu)^2 \simeq 2.3\times10^{-5}$). This results in a high degree of complexity in muon detection, due to the co-presence of a high level of noise arising from electron-positron pairs and bremsstrahlung photons. Several schemes have been numerically proposed to physically separate laser-driven muon beams from other secondary particles (see, e.g., Ref. \cite{Calvin2023}).

\edit{Here, we report on experimental evidence of the production of high-energy and directional muons. For the first time, the experimental setup allows selecting high-energy muons and provides a pathway for the separation of muons from leptonic and photon noise, validating a novel detection mechanism and recent numerical modelling. The detected muons are generated by the Bethe-Heitler process in a converter material irradiated by laser-wakefield accelerated GeV-scale electron beams. These muons exit the converter in a directional manner, with a divergence of the order of a 100 mrad.} Statistical analysis of the experimental data indicates detection of muon signal with a $(99.1\pm0.5)$\% confidence over noise, in close agreement with numerical simulations. Numerical extrapolation of these results to a PW-scale 10 Hz laser system such as EPAC in the UK \cite{EPAC} indicates that up to $10^4$ muons/s can be isolated and guided onto cm$^2$-scale areas. This configuration will allow, for instance, for the high resolution detection of strategically sensitive materials in high-Z thick containers within minutes \cite{Calvin2023}.\\

\noindent\textbf{II. EXPERIMENTAL SETUP}\\
The experiment was carried out at the Extreme Light Infrastructure - Nuclear Physics facility \cite{Gales2018} (Fig. \ref{Fig1}.a). 
The laser delivered 20 J on target with 28 fs duration and was focussed, using an f/26 off-axis parabola, to a peak intensity of $(6.0\pm1.2)\times10^{19}$ W/cm$^2$ at the edge of a 20 mm supersonic gas jet filled with a helium gas doped with 2\% nitrogen. 
A replenishable \unit[125]{$\mu$m}-thick kapton film was inserted after the jet to remove any residual laser light. 
A 45$^\circ$ wedged lead converter was inserted after the kapton film; by translating the wedge transversely to the laser axis, the converter thickness could be seamlessly varied between 0.5 and \unit[3]{cm}. Simulations indicate that the kapton film only marginally affects the spectral and spatial properties of the electron beam \cite{Raj2020}, and this is fully accounted for in the modelling of the experiment.

A \unit[10]{cm}-thick lead wall with a $3.5\times3.5$ cm$^2$ aperture on-axis was used to minimise off-axis noise arising from the electron beam interaction with the converter.
Two dipole magnets (Dipole 1 with a \unit[1]{T} field over \unit[30]{cm} and Dipole 2 with a \unit[1]{T} field over \unit[80]{cm}, see Fig.~\ref{Fig1}.a) were used to direct the muons onto the detection region, which was set up outside of the vacuum chamber. 
The muons exited the vacuum chamber through a \unit[2]{mm}-thick aluminium window. 
A Timepix3 \cite{Poikela2014,Darwish2020} detector was placed behind an L-shaped lead shielding configuration, consisting of a \unit[15]{cm}-thick vertical Pb wall and an horizontal \unit[20]{cm}-thick Pb slab (Fig. \ref{Fig1}.a). 
The Timepix3 detector used in the experiment consisted of a pixelated \unit[1]{mm}-thick silicon sensor which is flip chip bump-bonded to the Application-Specific Integrated Circuit (ASIC).
The sensor and ASIC are divided into a matrix of 256$\times$256 pixels of $\unit[55\times55]{\mu m^2}$, each of them measuring simultaneously the energy deposition and time of arrival. 
Katherine readout \cite{Burian2017} and TrackLab \cite{Manek2024,TrackLab} were used for detector readout, data acquisition and processing.

The electron beams accelerated in the plasma were first characterised using a LANEX scintillator screen placed after the first dipole magnet, without the converter. The electron beams exhibited a maximum energy exceeding 1 GeV, a divergence of the order of 1 mrad, and an overall charge above 200 MeV (lower limit of the magnetic spectrometer) of \unit[(560 $\pm$ 180)]{pC} (see Figs.~\ref{Fig1}.b and \ref{Fig1}.c). \\

\noindent\textbf{III. NUMERICAL MODELLING}\\
The average electron beam charge, spectrum (blue line in Fig.~\ref{Fig1}.c), and divergence were used as input for Monte-Carlo simulations using the code FLUKA \cite{Battistoni2015,Ahdida2022}. Five independent sets of simulations were performed, with 100 runs with $10^{10}$ primaries each to numerically extract the expected muon characteristics at source and at the detector plane, along with the spatial and spectral distribution of the noise expected at the detector plane. 
\begin{figure}[b!]
%\centering
\includegraphics[width=\linewidth]{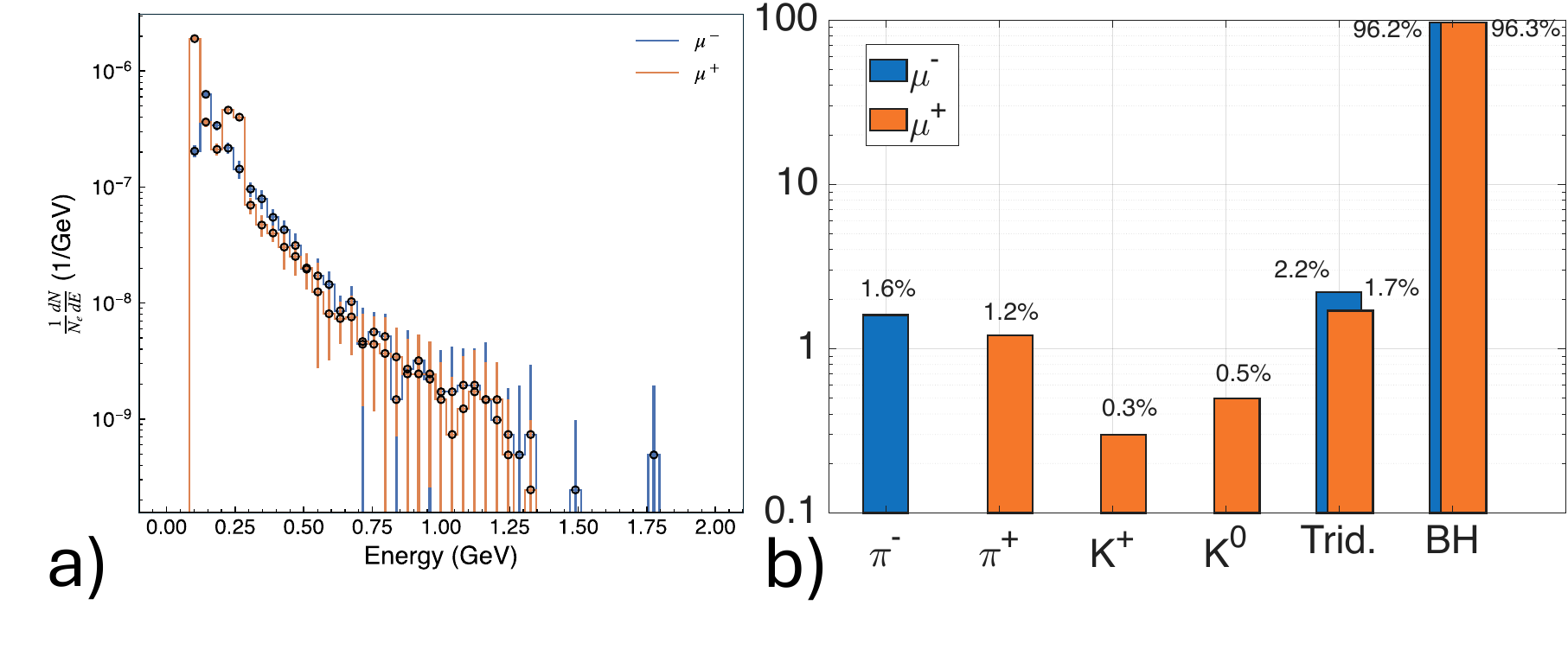}
\caption{\textbf{a.} Simulated spectrum of $\mu^-$ (orange) and $\mu^+$ (blue) generated during the interaction of an electron beam as depicted in Fig. \ref{Fig1} with a 2cm-thick lead converter. \textbf{b.} Distribution of parent particles for $\mu^-$ (blue) and $\mu^+$ (orange) reaching the detector. Trid. indicates direct electroproduction (trident), while BH denotes Bethe-Heitler pair production initiated by a bremsstrahlung photon in the converter.}
\label{Fig2}
\end{figure}
%A \unit[2]{cm}-thick Pb converter was used, since this is the thickness expected to maximise the flux of high-energy directional muons \cite{Calvin2023}. 
\edit{Extensive numerical modelling of muon production in a similar experimental configuration \cite{Calvin2023} indicates that the yield of high-energy directional muons increases linearly with target thickness up to approximately 2 cm. For thicker targets, the muon yield only marginally increases with the additional drawback of an enlarged divergence and, thus, decrease in flux. While optimum thickness will ultimately depend on the specific electron beam parameters and the muon population of interest, this thickness is optimum for this experimental configuration and is thus used hereafter.} 

Positive and negative muons present a similar spectrum (Fig. \ref{Fig2}.a) extending approximately up to 1.25 GeV and an energy-dependent divergence of the order of a hundred mrad (consistently with simulation results reported in Ref. \cite{Calvin2023}). \edit{Given the nature of the muon generation process, the muons exiting the converter target are expected to present a broadband, bremsstrahlung-like, spectrum virtually irrespective of the spectrum of the primary electron beam.}

Simulations indicate a total of $(0.79\pm0.02)\times10^{-7}$ $\mu^-$ and $(1.49\pm0.04)\times10^{-7}$ $\mu^+$ per primary electron. For an initial electron beam charge of $560\pm180$ pC, this translates into $280\pm90$ $\mu^-$ and $520\pm160$ $\mu^+$ escaping the converter target per shot. 

\begin{figure}[b!]
\centering
\includegraphics[width=1\linewidth]{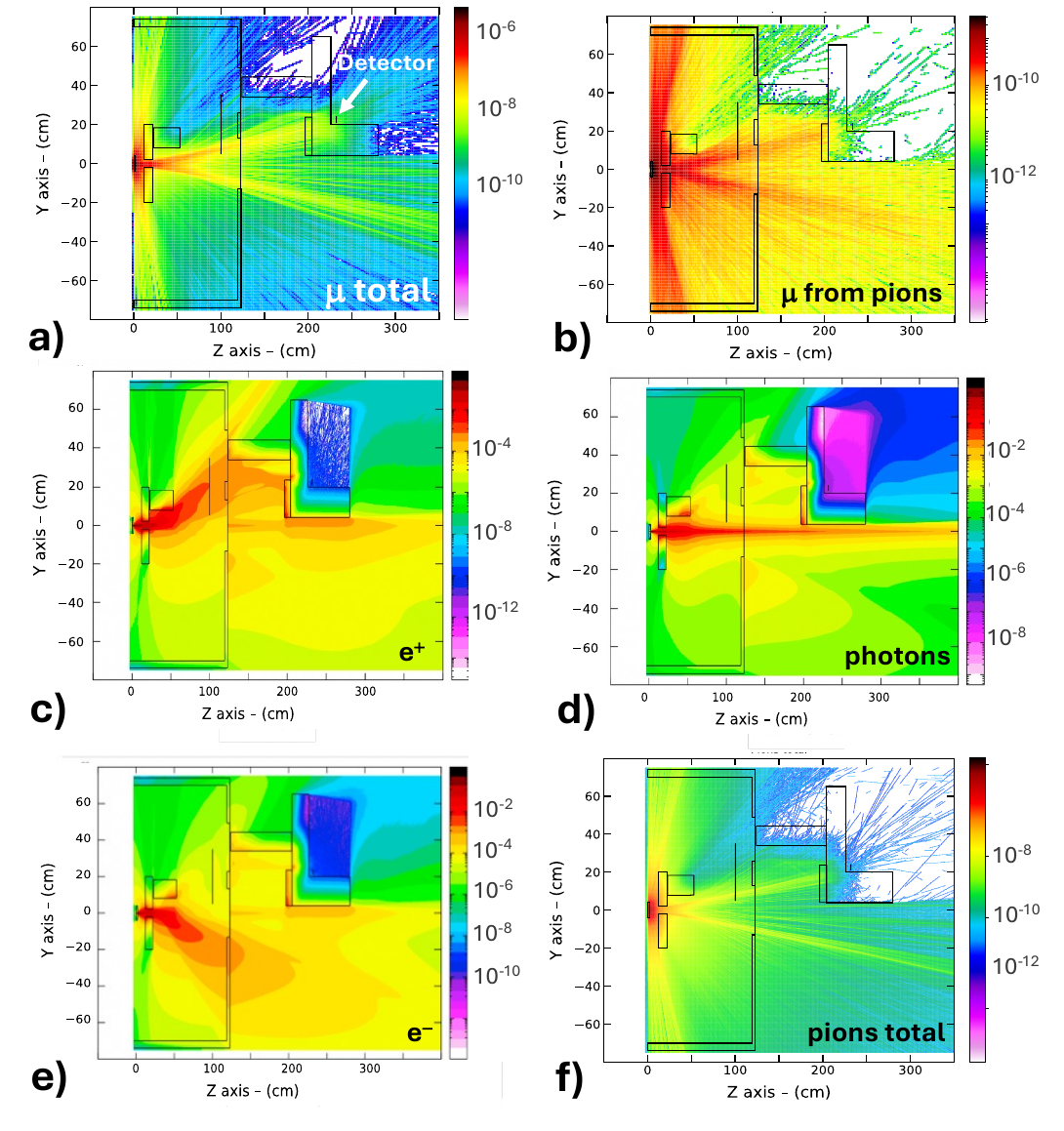}
\caption{Side-view of the simulated distribution through the experimental setup of all muons (a), muons generated by pion decay (b), positrons (c), photons (d), electrons (e), and all pions (f) through the experimental setup. The position of the detector is highlighted in frame (a) and all colorbars are in units of particle per primary electron per cm$^2$.}
\label{maps}
\end{figure}

\begin{figure}[t!]
\centering
\includegraphics[width=0.7\linewidth]{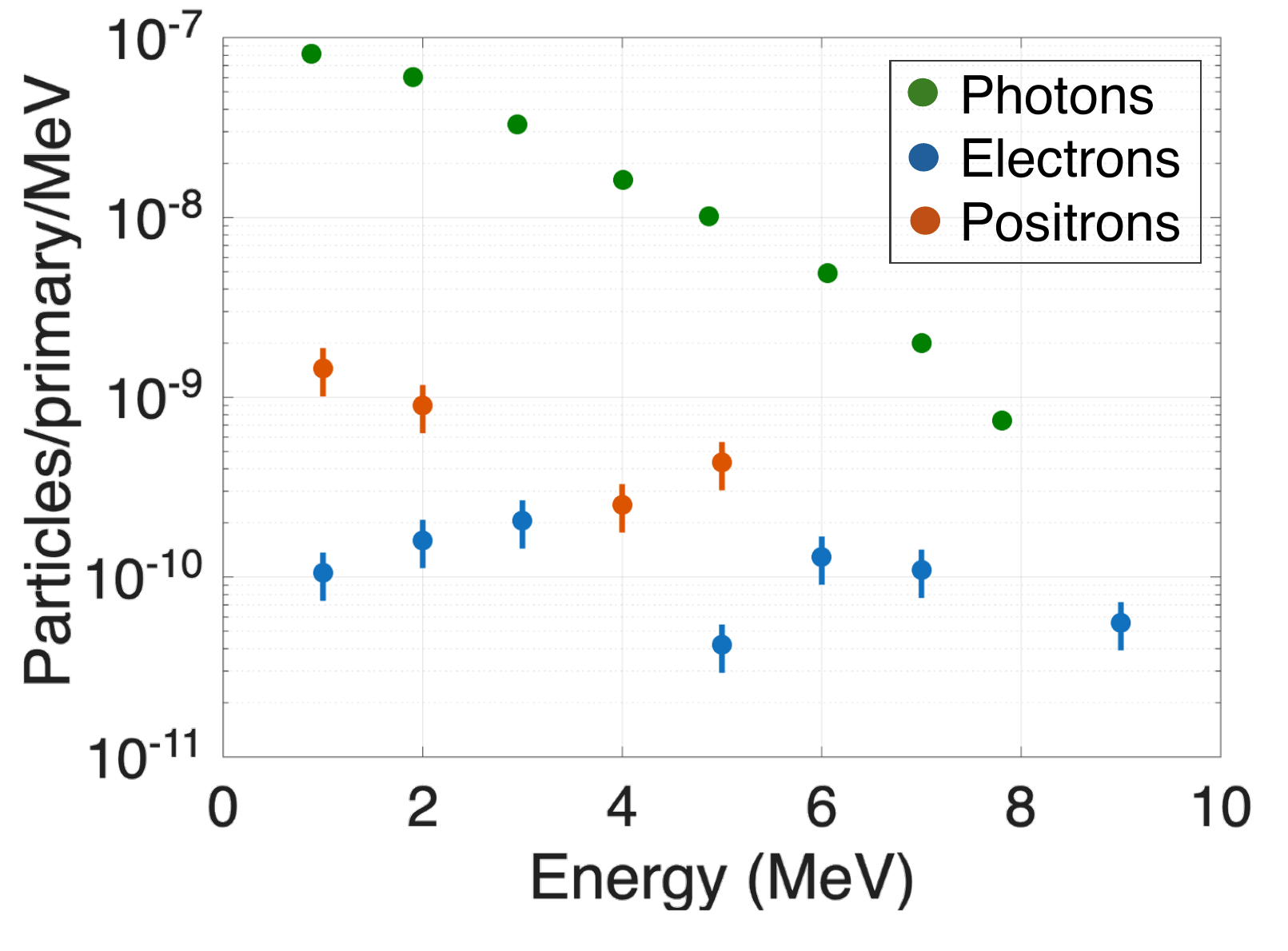}
\caption{\edit{Simulated spectra of photons, electrons, and positrons reaching the detector plane per primary electron per MeV.}}
\label{noise_spectra}
\end{figure}

A major obstacle in the detection of muons is the high-flux of scattered secondary particles exiting the converter. 
Numerical simulations indicate the generation of a total of $(5\pm3)\times10^8$ scattered electrons, bremsstrahlung photons, and generated positrons, with spectra extending up to the maximum energy of the primary electron beam (see also \cite{Sarri2015,Sarri_ncomm_2015}) and $>10^5$ charged pions with a broad angular distribution. In order to minimise the signal induced by these secondary particles in the detector, a specific shielding configuration has been designed and implemented in the experiment (Fig.~\ref{Fig1}.a). In this configuration, it is preferable to measure the positive muons, where the noise is lower. Monte-Carlo FLUKA simulations of the entire experimental configuration have been carried out to extract the number and spectral properties of muons and noise particles reaching the detector plane (1.4$\times$1.4 cm$^2$, as dictated by the size of the Timepix3 detector). The simulation results (see Fig. \ref{maps}) indicate approximately $(2.8\pm0.3)\times10^{-10}$ muons, $(8.0\pm0.4)\times10^{-10}$ electrons, $(3.0\pm0.1)\times10^{-9}$ positrons, $(2.0\pm0.1)\times10^{-7}$ gammas, and $(2.0\pm0.4)\times10^{-11}$ pions per primary electron reaching the detector per shot. The muons generated via pion decay represent a negligible fraction of the muons reaching the detector ($\approx2$\%, see Figs. \ref{Fig2}.b and \ref{maps}.b). Due to the magnetic transport line, the muons reaching the detector plane have an energy of \unit[($300\pm50$)]{MeV}; the other noise particles \edit{reaching the detector} (i.e., electrons, positrons, and gammas) have instead a broad energy spectrum up to approximately 10 MeV \edit{(Fig \ref{noise_spectra})}. %The uncertainty in actual particle yields is dominated by the shot-to-shot fluctuations in the spectrum and charge of the primary electron beam ($\simeq 30\%$, see Fig. \ref{Fig1}.c). 

The muon signal can be statistically distinguished from other particles by the characteristics of energy deposition in the detector, a standard technique in particle physics experiments (see, e.g., Ref. \cite{ATLAS_muons}). Although methods based on measuring muon lifetime are also possible \edit{\cite{Terzani2024,Zhang_muons}}, the high electromagnetic noise generated in the chamber by the high-power laser in this experiment prevented a clear signal identification using this technique. High-energy muons produce distinctive track shapes and energy deposition patterns on the detector, distinguishing them from MeV-scale leptonic noise (electrons hereafter for definitiveness), ions, and neutrons. While pions could also in principle contribute to the signal \cite{Schumaker_2018}, the simulation results discussed above indicate that only 0.07$\pm$0.02 pions reach the detector per shot. This low likelihood arises from their wide-angle emission combined with the magnetic transport line and the lead shielding in front of the detector (see Fig. \ref{maps}.f). Our simulations also confirmed that Timepix3 based on thin silicon detectors are also inefficient in detecting MeV-scale photons, with negligible energy deposition (see also Ref. \cite{itoh}). 

\begin{figure}[t!]
\centering
\includegraphics[width=1\linewidth]{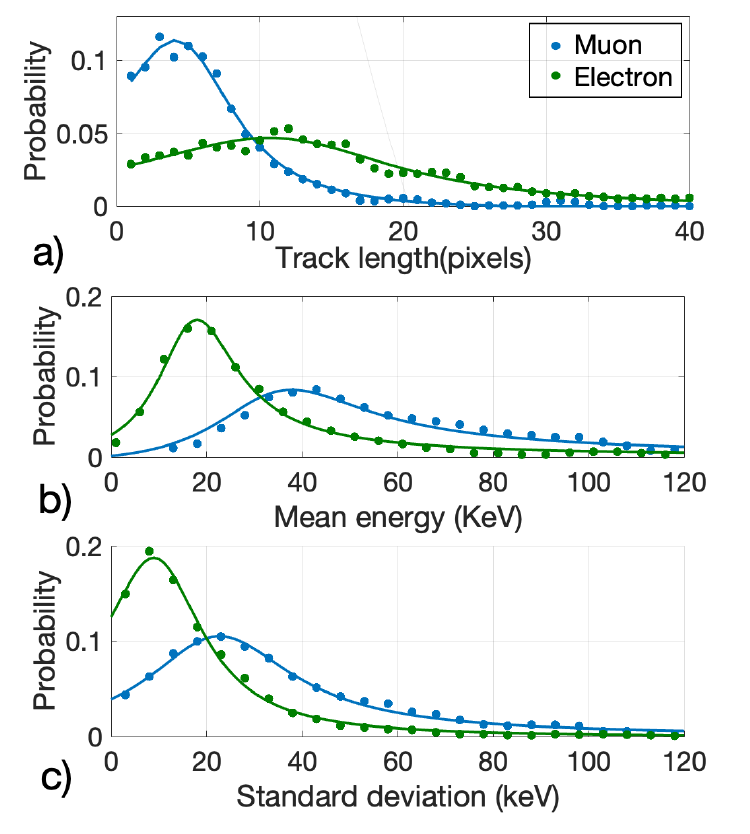}
\caption{Simulated distribution of track length (a), mean energy deposited (b), and standard deviation within the track (c) from FLUKA simulations for $300\pm50$ MeV muons (blue) and electrons with a flat energy distribution extending up to 10 MeV (green) interacting with the TimePix3 detector (simulation details in the article). The simulated data points have been fitted with a rational function of the form: $y(x) = (p_1 x + p_2)/(x^2 + q_1 x + q_2)$.}
\label{fig:tracks}
\end{figure}

For the other noise particles, a series of simulations were performed using the Monte-Carlo code FLUKA of the response expected on the detector for $300\pm50$ MeV muons, MeV-scale electrons, neutrons, and protons. In this set of simulations, a \unit[1]{mm}-thick Si slab with $256\times256$ pixels with a size of \unit[55]{$\mu$m} $\times$ \unit[55]{$\mu$m} was irradiated $10^6$ different times with each particle, with electrons and positrons (generally referred as electrons hereafter) having an energy randomly sampled from a uniform distribution from zero to 10 MeV. 
Ions and neutrons are expected to arrive at the detector plane significantly after the temporal gating applied to the detector (discussed below) and can thus be easily discarded by time-filtering the signal. For muons and electrons, the distributions of the mean and standard deviation of the energy deposited per track and the length of the tracks are shown in Fig. \ref{fig:tracks}. In all cases, Kolmogorov-Smirnov tests of the distributions indicated a $p$-value < 0.001 confirming that the distributions are statistically different. Muon tracks are generally shorter, with a higher mean and standard deviation of the energy deposited. This is accompanied by a different shape of the tracks; muon tracks have negligible curvature and are never observed to be more than 1 pixel wide. \edit{These numerical results are consistent with the significance difference in energy deposition characteristics between high-energy muons and low energy electrons or positrons, with the latter experiencing larger scattering angles and a more continuous energy transfer throughout the material.}\\

\begin{figure}[b!]
\centering
\includegraphics[width=1\linewidth]{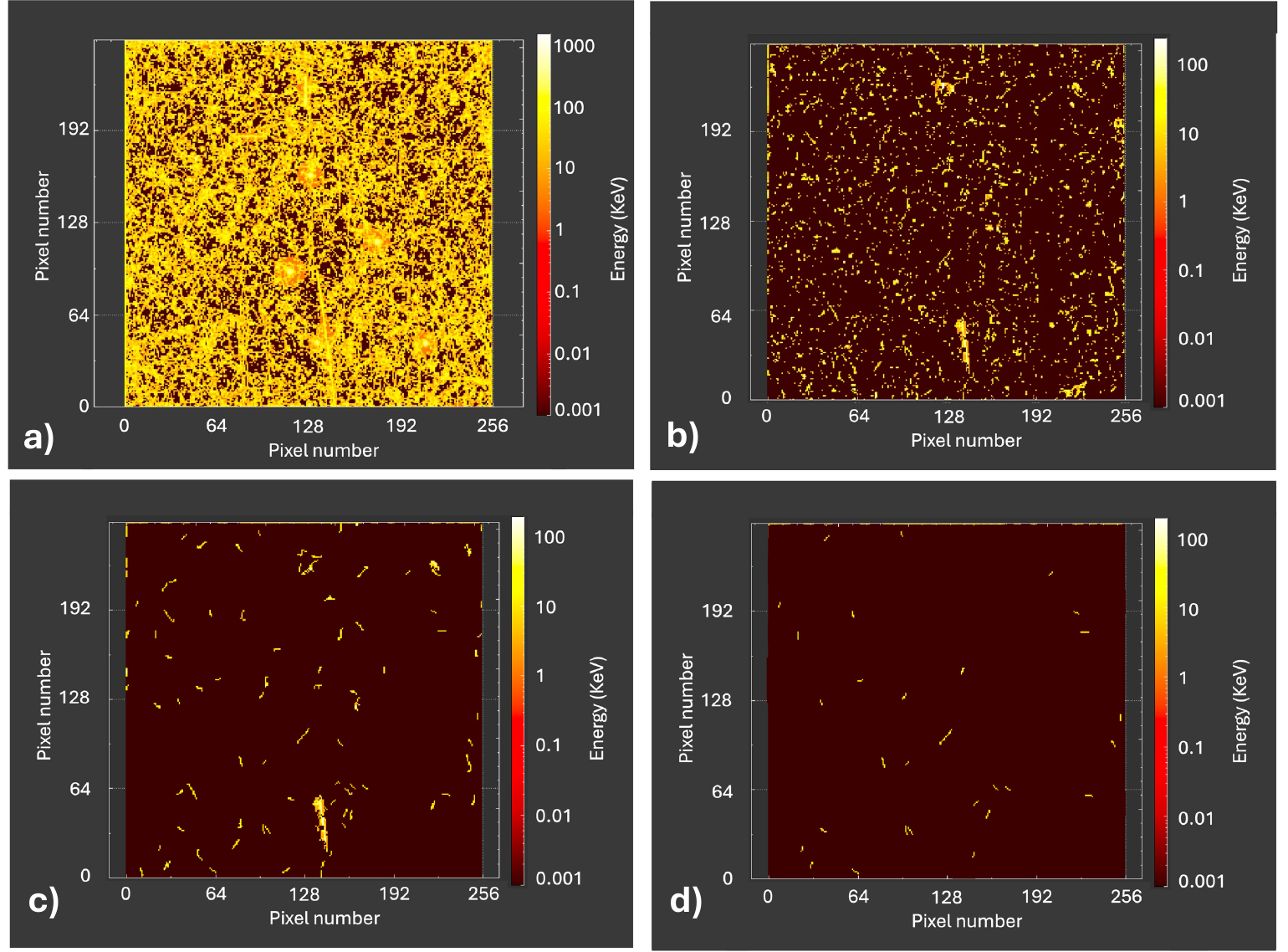}
\caption{a) TimePix3 raw data accumulated over 10 consecutive shots with a 2 cm Pb converter. b) TimePix3 data from (a) filtered with a 2ns temporal window around the expected muon arrival time (details in the article). c) TimePix3 data from (b) with the additional filter of track straightness. d) TimePix3 data from (c) with only 1 pixel-wide tracks.}
\label{TimePix3}
\end{figure}

\noindent\textbf{IV. EXPERIMENTAL RESULTS AND ANALYSIS}\\
The raw data recorded by the Timepix3 detector for a series of 10 consecutive shots with a \unit[2]{cm} converter target is shown in Fig. \ref{TimePix3}.a. 

The raw data were filtered first by only selecting tracks that arrived at the detector within $8\pm2$ ns after the laser arrival time on the gas-jet, as this is the expected temporal window of arrival of the muons at the detector plane. Ion and neutron signal is already effectively removed by this filter, as these particles reach the detector plane at least 11 ns after the interaction (Fig. \ref{TimePix3}.b).
The raw data is then further filtered by selecting only tracks that do not present a significant curvature (Fig. \ref{TimePix3}.c) and that are only 1 pixel wide (Fig. \ref{TimePix3}.d). 
\begin{figure*}[b!]
\centering
\includegraphics[width=1\linewidth]{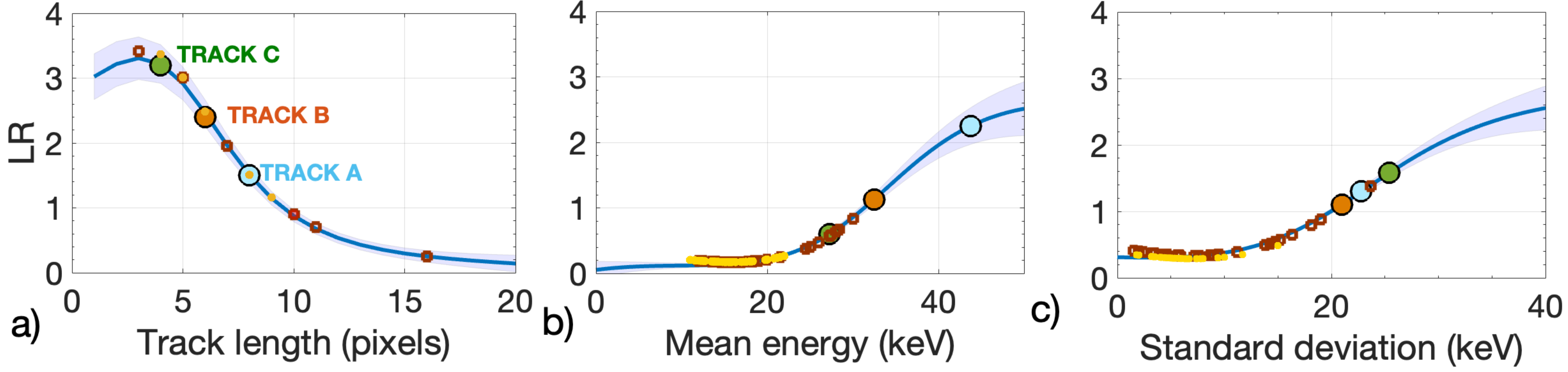}
\caption{Likelihood ratios (blue solid lines) and associated uncertainty (blue shaded regions) for track length (\textbf{a.}), mean energy deposited (\textbf{b.}), and standard deviation of energy deposited per pixel across the track (\textbf{c.}). Brown squares and yellow dots indicate the length, mean energy, and standard deviation for every track for the converter thicknesses used in this study (2 cm for the brown squares and 1 and 1.5 cm for the yellow dots). The labelled large circles highlight the three tracks with the highest combined likelihood ratio (4.4 for track A, 3.0 for track B, and 3.0 for track C)}. 
\label{Fig4}
\end{figure*}

\begin{figure}[t!]
\centering
\includegraphics[width=1\linewidth]{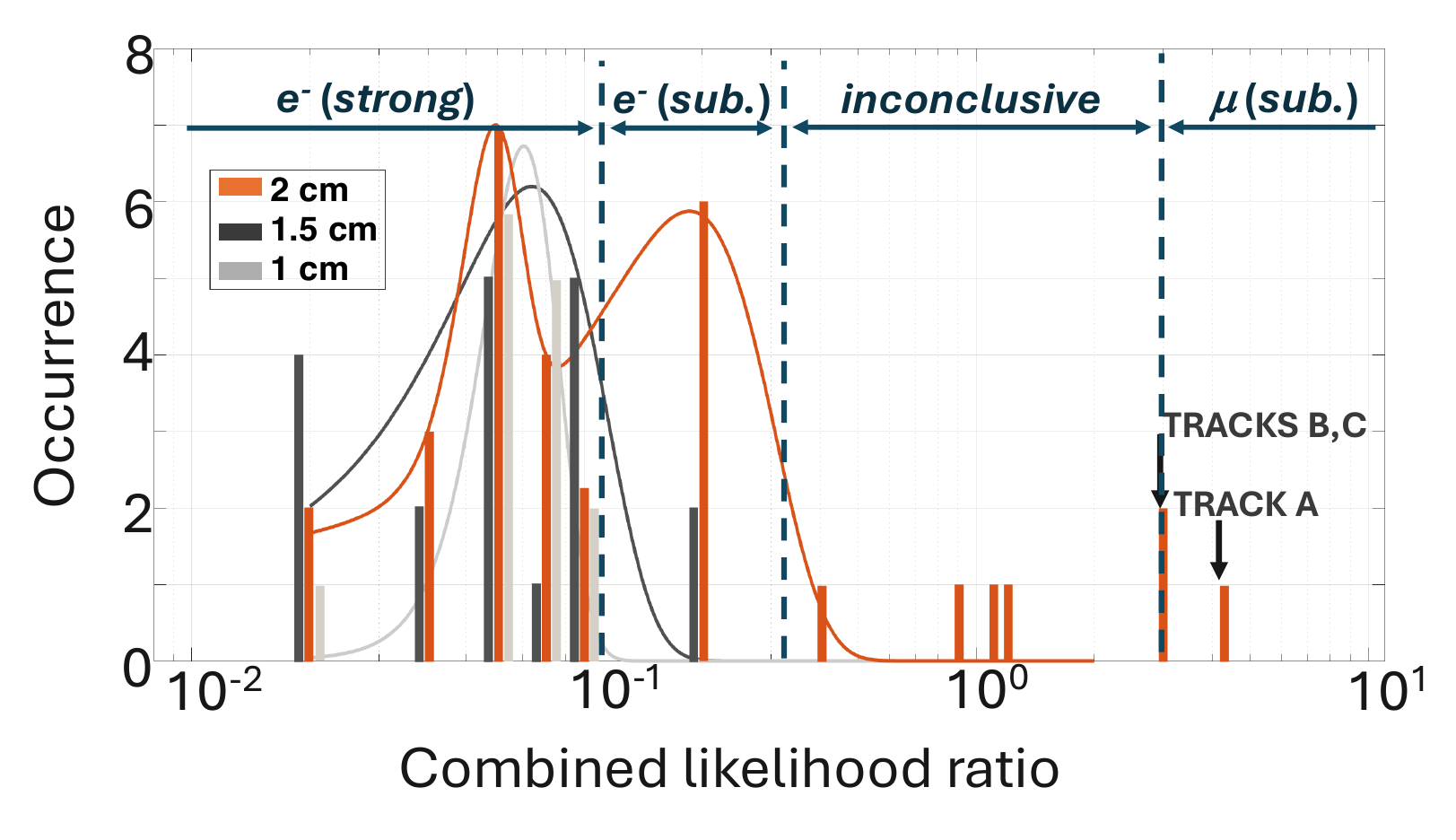}
\caption{Histogram of combined likelihood ratios for the tracks recorded with the 1 cm (grey), 1.5 cm (black), and 2 cm (orange) converters. Gaussian fits are included as a guide for the eye.}
\label{Fig5}
\end{figure}

The filtered data still show approximately 30 tracks, which could in principle be produced by either a muon or an electron. In order to further distinguish between electron and muon signal, likelihood ratios (LRs) for the length of the tracks and the mean and standard deviation of the energy deposited are shown in Fig.~\ref{Fig4}, with brown dots representing each track in Fig. \ref{TimePix3}.d. 
\edit{The likelihood ratio that a track has a property $Q$ (for example, length) given that it is generated by a particle type $p$ and not any other particle ($\neg p$) is defined as $\mathrm{LR}[Q\lvert p] \equiv P[Q\lvert p] / P[Q\lvert \neg p]$, where the probabilities $P$ are extracted from the distributions shown in Fig.~\ref{fig:tracks}.
A track can be parameterised by its length $\ell$, the mean energy deposited $\bar{E}$, and its standard deviation $\sigma_E$. The combined likelihood for a given track $T_i$ is then the product $\mathrm{LR}[T_i\vert p] = \mathrm{LR}[\ell_i\lvert p] \cdot \mathrm{LR}[\bar{E}_i\lvert p] \cdot \mathrm{LR}[\sigma_{E,i}\lvert p]$.
}
The corresponding histogram of the combined LR for each track in Fig. \ref{TimePix3}.d is shown in orange in Fig. \ref{Fig5}. An LR above 3 (below 0.3) is considered substantial evidence of muon (electron) behaviour, while an LR above 10 (below 0.1) is considered strong evidence \cite{Jeffreys_1961}. The histogram is well approximated by a two-Gaussian distribution with means at $0.06\pm0.01$ and $0.18\pm0.02$, providing strong preliminary indication that most tracks are produced by electrons. Notably, three tracks (labelled A,B, and C in Figs. \ref{Fig4} and \ref{Fig5}) have combined LRs above 3 ($4.4\pm0.2$ and $3.0\pm0.1$), first preliminary indication of being produced by a muon.

For a more quantitative analysis, the LRs have been used to extract the probability that each track in Fig. \ref{TimePix3} is produced by a muon or an electron.
For this, the pre-test probability has been calculated by applying the same filters of time of arrival, straightness, and track width to the simulation data. 
After this process, simulations indicate $N_e=25\pm2$ electron tracks and $N_\mu=9\pm1$ muon tracks over ten shots, yielding a pre-test probability \edit{
$P[\mu] = N_\mu/(N_e+N_\mu) = (26.5\pm3.6)$\%, and pre-test odds $\mathcal{O}[\mu]=P[\mu]/(1-P[\mu]) = 0.36\pm 0.05$ for a track to be a muon. The equivalent pre-test probability and odds for a track to not be a muon, i.e. is an electron, are $P[\neg \mu] = P[e] = (73.5 \pm 10.1)\%$ and $\mathcal{O}[\neg\mu] = \mathcal{O}[e] = 2.77\pm 0.53$ respectively.
The post-test odds that a given track, $T_i$, is of particle type $p$ can thus be calculated as: $\mathcal{O}[T_i\lvert p] = \mathrm{LR}[T_i \lvert p] \times \mathcal{O}[p]$.
The probability that each track has been generated by a muon can thus be expressed as $P[\mu \lvert T_i] = \mathcal{O}[T_i \lvert \mu]/(1+\mathcal{O}[T_i\lvert \mu])$, with the probability of it being generated by an electron being necessarily $P[e\lvert T_i] = 1- P[\mu\lvert T_i]$. Given the fact that each track can be treated independently, the probability that at least one muon has been detected can thus be calculated as:}

\begin{equation}
P[\mu \geq 1] 
= 1 - \displaystyle \prod_{i=1}^{n} P[\neg \mu\lvert T_i]
=1-\displaystyle \prod_{i=1}^{n} \left(1-\frac{\mathcal{O}[T_i\lvert \mu]}{1+\mathcal{O}[T_i\lvert \mu]}\right).
\end{equation}

For all the tracks shown in Fig. \ref{TimePix3}.d, the probability of having detected at least one muon is $P[\mu \geq 1] = (99.1\pm 0.5)$\%.

To further confirm this evidence, additional shots were taken in the same conditions but with a thinner converter target: 5 shots at \unit[1]{cm} and 5 shots at \unit[1.5]{cm}.  Simulations with these converter thicknesses show no muons reaching the detector over 5 independent runs, indicating that the number of muons per primary electron potentially reaching the detector per shot is below $10^{-12}$. 
For an electron beam charge of \unit[($560\pm180$)]{pC}, one would then expect less than 0.03 muons to reach the detector over ten shots. Applying the same filtering and the analysis discussed above, the 1 and 1.5 cm shots result in the histograms of combined LRs shown in Fig.~\ref{Fig5} (single LRs for each track also overplotted in Fig. \ref{Fig4} as yellow dots). 
The histograms are well approximated by Gaussian distributions with a mean of 0.070$\pm$0.005 and a standard deviation of 0.11$\pm$0.05 and $0.015\pm0.008$ for the two target thicknesses, strong indication that the tracks are all produced by leptonic noise, as expected. For these thinner targets, the total probability of having detected a muon is $P[\mu \geq 1] < 14$\%.

\noindent\textbf{V. CONCLUSIONS}\\
We report on the generation of directional and high-energy muons from a laser-wakefield accelerator driven by a PW-class laser. Data analysis indicates a $(99.1\pm0.5)$\% confidence of muon detection over noise, in excellent agreement with numerical modelling.

The results reported here can be readily scaled to laser systems of high peak power or repetition rate. %Numerical modelling indicates approximately 100 muons generated per shot for a 500 pC, GeV-scale electron beam. 
Extrapolation to a nC-scale 2 GeV electron beam, as attainable by PW-scale lasers \cite{Poder_2024}, indicates that up to $10^3$ muons per shot can be isolated and guided by a suitable transport line \cite{Calvin2023}. A 10 Hz PW laser such as EPAC in the UK \cite{EPAC} would then be able to produce and guide up to $10^4$ muons/s in shielded cm$^2$-scale areas, enabling radiography of extended high-Z objects within tens of seconds \cite{Calvin2023}.
10 PW-scale laser facilities are also now starting to enter the international landscape \cite{Danson2019} (see, e.g., Vulcan 2020 \cite{Vulcan}, OPAL \cite{OPAL}, and ELI-NP \cite{Gales2018}). 
Numerical simulations show that the 10PW line at ELI-NP can produce multi-nC electron beams with multi-GeV maximum energies \cite{Calvin2023}. The interaction of such electron beams with a 2 cm converter is predicted to produce more than $10^5$ muons with a broadband energy distribution peaking at 0.7 GeV. \edit{The increase in muon yield expected with PW-class laser systems can also be accompanied by a more complex transport line and shielding configuration as proposed and numerically studied in \cite{Calvin2023}. Adopting an experimental configuration of that kind can thus conservatively direct up to $10^4$ muons in a relatively background-free region with a 10x10 cm$^2$ area \cite{Calvin2023}}. Even operating at one shot per minute, 10 PW-scale lasers would enable muon radiography of extended high-Z materials to be carried out within a few minutes \cite{Calvin2023}. 

\textbf{Acknowledgments}: G.S. wishes to acknowledge support from EPSRC (grant No. EP/V044397/1, EP/V049186/1, and 2934194) 
and EU Horizon Europe R\&I Grant No. 101079773 (EuPRAXIA-PP).  P.M. wishes to acknowledge support from EPSRC (grant No. EP/V049232/1). B.B. and P.S. acknowledge funding from the Czech Science Foundation under Registration Number GM23-04869M.
R.E.M. has been supported by the Global Postdoc Fellowship Program of the Czech Technical University in Prague. The authors wish to acknowledge support from the Vulcan dark period community support programme 23-1 from the Rutherford Appleton Laboratory.
This work was supported by the Extreme Light Infrastructure–Nuclear Physics (ELI-NP) Phase II, a project co-financed by the Romanian Government and the European Union through the European Regional Development Fund, by the Romanian Ministry of Education and Research CNCS-UEFISCDI (Project No. PN-IIIP4-IDPCCF-2016-0164) and Nucleu Projects Grant No. PN 23210105. The Romanian government also supports ELI-NP through IOSIN funds as a Facility of National Interest.

\bibliography{Refs}

\end{document}